\documentclass[aps,pra,superscriptaddress,preprint,showpacs]{revtex4-1}
\usepackage{graphicx}
\usepackage{amsmath}
\usepackage{amsfonts}
\usepackage{epstopdf}
\usepackage{euscript}
\usepackage{subfigure}
\usepackage[usenames]{color}

\begin{document}
\title{Ion-momentum imaging of dissociative electron attachment dynamics in acetylene}
\author{M.~Fogle}
\affiliation{Department of Physics, Auburn University, Auburn, Alabama 36849, USA}
\author{D.~J.~Haxton}
\affiliation{Lawrence Berkeley National Laboratory, Chemical Sciences, Berkeley, California 94720, USA}
\author{A.~L.~Landers}
\affiliation{Department of Physics, Auburn University, Auburn, Alabama 36849, USA}
\author{A.~E.~Orel}
\affiliation{Department of Chemical Engineering and Materials Science, University of California, Davis, California 95616, USA}
\author{T.~N.~Rescigno}
\affiliation{Lawrence Berkeley National Laboratory, Chemical Sciences, Berkeley, California 94720, USA}

\date{\today}


\begin{abstract}
We present experimental results for dissociative electron attachment to acetylene near the 3~eV $^2\Pi_g$ resonance. In particular, we use an ion-momentum imaging technique to investigate the dissociation channel leading to C$_2$H$^-$ fragments. From our measured ion-momentum results we extract fragment kinetic energy and angular distributions. We directly observe a significant dissociation bending dynamic associated with the formation of the transitory negative ion. In modeling this bending dynamic with \emph{ab initio} electronic structure and fixed-nuclei scattering calculations we obtain good agreement with the experiment.  

\end{abstract}

\maketitle

\section{Introduction}
The acetylene molecule represents an ideal system to investigate polyatomic dissociation dynamics associated with dissociative electron attachment (DEA).  It is the smallest unsaturated hydrocarbon and is linear at equilibrium. There is an unoccupied $\pi$ orbital making it a prototype to investigate the dynamics associated to $\pi^*$-shape resonances.

Electron-driven chemistry of acetylene is important in key semiconductor and nanotechnology applications, e.g, acetylene is commonly used for carbon film deposition \cite{lettington1998, ashfold2001}. It is also of interest in astrophysics, e.g., acetylene has been identified in Titan's atmosphere \cite{hanel1981, abbys2002, vuitton2009, suits2009}.

Acetylene has drawn a considerable amount of theoretical and experimental investigation, however most of those studies have not been aimed at understanding transitory anion formation and its associated dissociation dynamics.  Early DEA experiments to acetylene revealed a resonance near 3~eV that yielded C$_2$H$^-$ fragments. This was attributed to a $\pi^*$-shape resonance. A weak band of resonances between 6--8~eV was also found to yield C$_2^-$ fragments \cite{trepka1963, azria1972}.  Vibrational excitation measurements of Andric and Hall \cite{andric1988} revealed a $\pi^*$ resonance at 2.6~eV and a $\sigma^*$ resonance at 6.2~eV. Kochem et al. \cite{kochem1985} studied electron-impact vibrational excitation below 4~eV and attributed the $^2\Pi_g$ temporary anion state to the resonance at 2.6~eV. Gauf et al. \cite{gauf2013} also studied elastic electron scattering from acetylene and found the $^2\Pi_g$ resonance near 3~eV to contribute significantly to the differential scattering cross section. Dressler and Allan \cite{dressler1987} performed electron transmission experiments on acetylene and found that predominantly C$_2$H$^-$ fragments resulted at the 3~eV $^2\Pi_g$ resonance with a nonvertical onset at 2.6~eV. May et al. \cite{may2008, may2009} have made recent absolute measurements of DEA to acetylene, and its deuterated isotopolog, using an adapted total-ion collection tube with time-of-flight capabilities for mass resolving power. They observe a peak in the $^2\Pi_g$ resonance at 2.95~eV. Syzma\'nska et al. \cite{szymanska2014} have also recently measured anion production from acetylene via DEA and dipolar dissociation. Their results for DEA resonance positions are in agreement with previous experiments.

 Chourou and Orel \cite{Chourou} made \emph{ab initio} calculations of DEA to acetylene and the expected three-dimensional bending dynamics necessary to describe the dissociation process. Several of the previous experimental studies using electron transmission and scattering techniques inferred a bending dynamic attached to the $^2\Pi_g$ transitory negative ion state; however this had not been observed directly with any detail. Syzma\'nska et al. made velocity slice images of the anion fragments formed due to DEA but no detailed angular structure or kinetic energy information was obtained due to poor resolution. In this paper, we present ion-momentum imaging of the C$_2$H$^-$ fragments formed via the 3~eV $^2\Pi_g$ DEA resonance and a theoretical model to account for the nonaxial recoil dynamics associated with bending during the dissociation process.

\section{Experiment}
The DEA measurements of this study were made using the momentum imaging apparatus at Auburn University, which has been discussed previously in more detail \cite{moradmand13b}. The target molecular beam was made by expanding acetylene gas adiabatically through a 10~$\mu$m nozzle to form a supersonic gas jet. A 300~$\mu$m skimmer then selected the central portion of this jet to form a molecular beam that was passed into the interaction region. The molecular beam had a width of approximately 2~mm at the interaction point with the crossed electron beam.

\begin{figure}[t]
\includegraphics[scale=0.6]{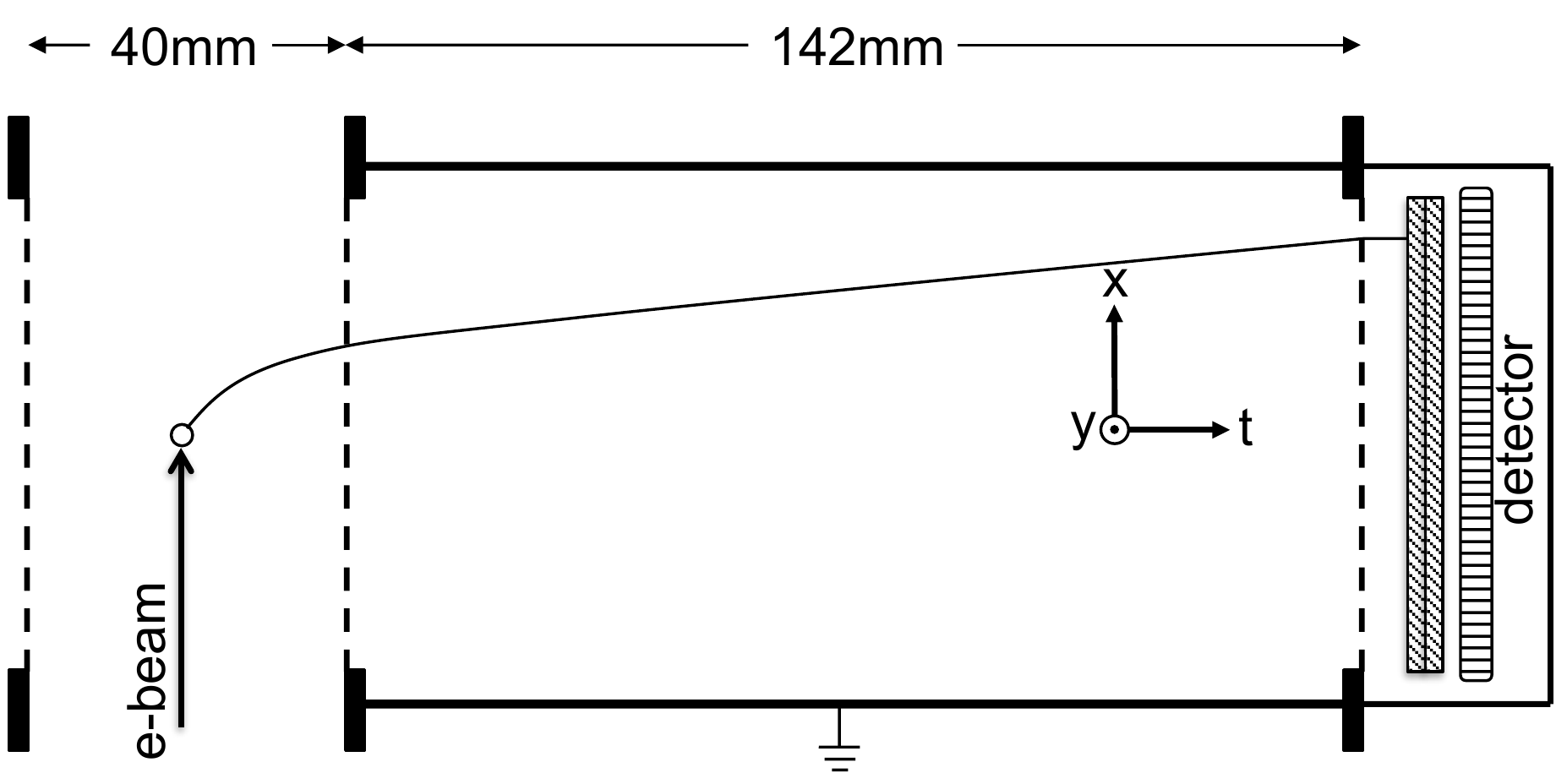}
\caption{Schematic of the anion fragment spectrometer. The small circle represents the molecular beam target propagating out of the page and crossed by the electron beam pulse. The vertical dashed lines represent 90\% transmission stainless steel grids. The far left grid is a pusher that is pulsed from ground to a negative potential for anion extraction from the interaction region. The $x,y,t$ coordinate axes of the system are also shown.}
\label{spect}
\end{figure}

Figure \ref{spect} shows a schematic of the electrostatic spectrometer of the apparatus. The interaction region is situated within the spectrometer as shown in Fig.~\ref{spect}. At the interaction region the molecular acetylene beam was crossed orthogonally in an electrostatically field-free region by a 200-ns duration, 20 kHz pulsed electron beam of $3.0\pm 0.5$~eV. After each electron pulse, a pusher grid was pulsed for a duration of 1~$\mu$s from ground to -40~V after a 300-ns delay. This provided an impulse to anions formed during dissociation and resulted in complete $4\pi$ collection of the C$_2$H$^-$ anion fragments. Anion fragments were recorded by a chevron configuration of 80-mm microchannel plates with a delay-line anode for position determination. The time-of-flight was also recorded between the electron pulse and the particle impact on the detector. The C$_2$H$^-$ anions had flight times of approximately 40~$\mu$s. The position and timing information allow for a complete determination of the initial momentum vector of a C$_2$H$^-$ anion fragment upon dissociation. Recorded dissociation event data are analyzed off line by defining a dissociation sphere through which slicing planes can be selected to observe angular fragment distributions and associated kinetic energy release.

\begin{figure}[t]
\includegraphics[scale=0.6]{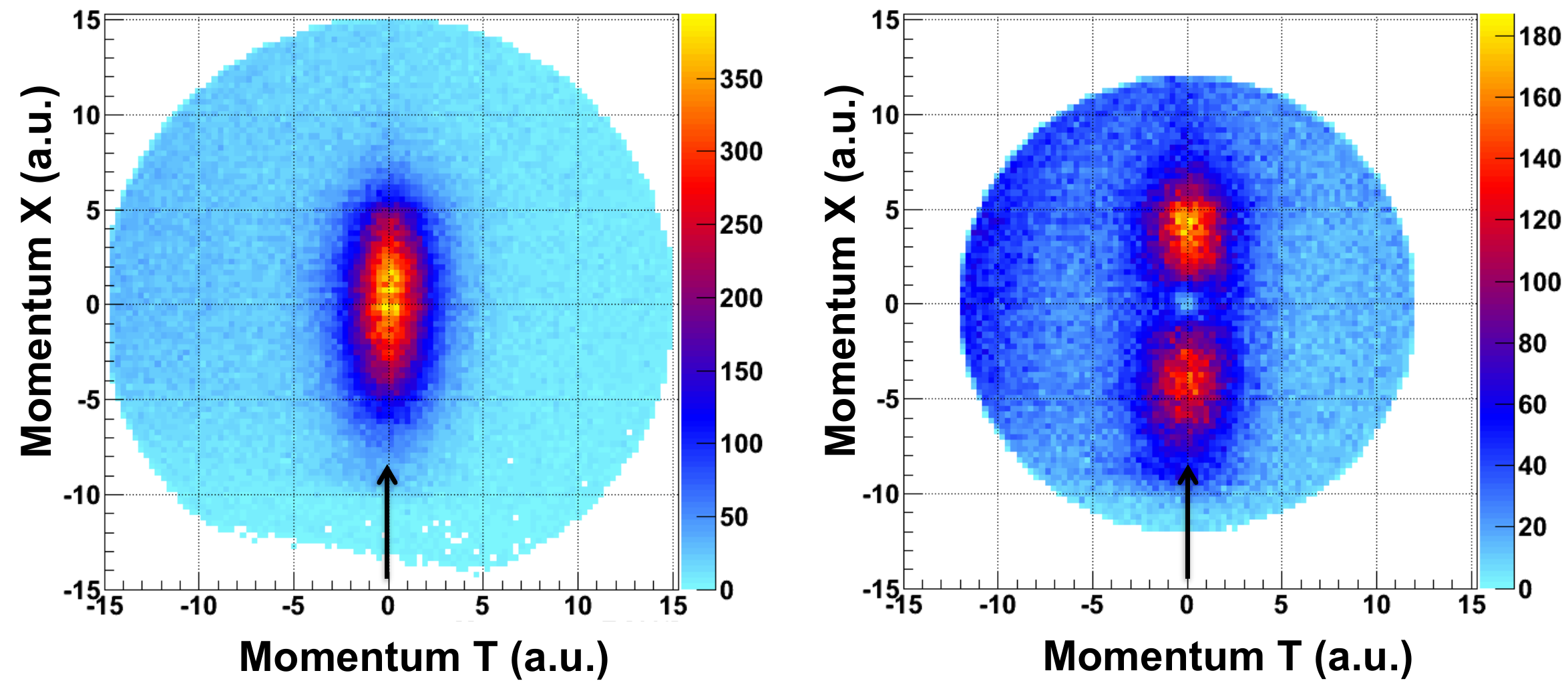}
\caption{(color online) (left) An $x,t$ momentum-space slice of the C$_2$H$^-$ dissociation sphere. (right) The same momentum-space data weighted for solid angle (see text for details). The incoming electron direction is indicated by the arrow.}
\label{xtplot}
\end{figure}

As can be seen in the left pane of Fig.~\ref{xtplot}, such a momentum sphere slice in the $x,t$-plane is presented to show the C$_2$H$^-$ fragment distributions in momentum space. It is important to note that such a slice of the momentum sphere needs to be appropriately weighted for solid angle, otherwise particles with low kinetic energy release would be over-emphasized. The right pane of Fig.~\ref{xtplot} is the same data weighted to account for the changing solid angle by constraining the elevation angle, $\phi=\tan^{-1}\left( P_Y/\sqrt{P_X^2+P_T^2} \right)$, of the momentum to within $\pm$5$^\circ$. This effectively forms a wedge whose vertex is at the origin of the momentum space and is revolved around the selected slicing plane. The constraining angle of the wedge is chosen to coincide with the estimated momentum resolution of the experiment. As the radius increases, the angle of acceptance is held constant so that fragments of differing energy are treated equally. 

\section{Theory}
We have carried out {\em ab initio} electronic structure and fixed-nuclei scattering calculations to connect the experimentally observed angular distributions to molecular-frame dissociation dynamics. Our approach is based on evaluating the quantum mechanical entrance amplitude \cite{Omalley66}, whose squared modulus gives the electron attachment probability in the body frame of the molecular target. The entrance amplitude can in turn be evaluated from an analysis of the fixed-nuclei S-matrix~\cite{Haxton06} which we calculate using the complex Kohn variational method~\cite{Rescigno95}. 

The Kohn calculations were carried out using the same augmented triple-zeta-plus-polarization basis employed by Chourou and Orel~\cite{Chourou}. The neutral N-electron target was described by a 106-term complete active space (CAS) configuration-interaction wave function using nine molecular orbitals (seven occupied self-consistent field plus a 2 improved virtual $\pi$* orbitals), calculated with the constraint that the two carbon 1s molecular orbitals remain doubly occupied. The acetylene molecule is linear at equilibrium geometry and belongs to the D$_{\infty h}$ point group. However, all our calculations were carried out in C$_s$ symmetry to maintain consistency between equilibrium and asymmetric, bent geometries. The (N+1)-electron Kohn trial function included the open-channel target ground-state.  In addition, the additional 105 A$'$ and 90 A$''$ singlet states that could be constructed from the CAS orbitals were included as closed-channels. We also included all of the (N+1)-electron ``penetration" terms~\cite{Rescigno95} that can be formed from the CAS orbitals.  To maintain a balance between the N- and (N+1)-electron wave functions, we did not include any triplet target states in the expansion, which we found to over-correlate the negative ion resonance relative to the neutral target. The Kohn trial function also included regular and outgoing numerical continuum function up to $l,|m|$=4. This prescription gave a trial function with $\sim$8000 terms. At equilibrium geometry, we find a $^2\Pi_g$(A$'$) resonance at $\sim$3 eV with a total width of 1.75 eV, in good agreement with the earlier results of Chourou and Orel~\cite{Chourou}.

The attachment probability can, in principle, be calculated using formal resonance theory by partitioning the full scattering wave function into resonance and background parts and defining the entrance amplitude as the matrix element of the electronic Hamiltonian between the resonance and background wave functions. 

\begin{figure}
\includegraphics[scale=0.8]{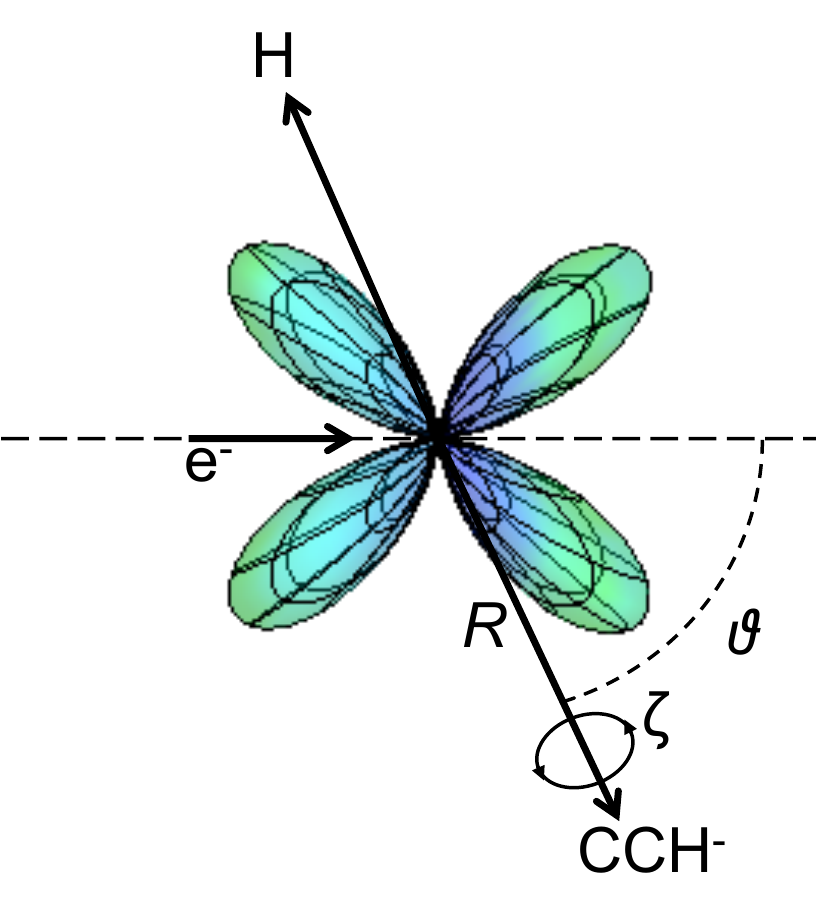}
\caption{(color online) Schematic of the modified axial recoil approximation used in the calculations. The scattering angle $\theta$ is the angle between the incident electron direction and the recoil axis {\bf R}. $\zeta$ is the angle azimuthal to {\bf R}. To simulate post-attachment bending, the recoil axis is first rotated while holding the attachment probability, shown as the shaded lobed region, fixed and then averaged about $\zeta$ to produce an angular distribution in $\theta$.}
\label{thetafig}
\end{figure}
A more practical approach is to evaluate the entrance amplitude in terms of quantities obtained from an analysis of the calculated fixed-nuclei $S$ matrix, as outlined in Ref.~\cite{Haxton06}. In that work, we showed that the entrance amplitude can be expressed as
\begin{equation}
V_a=\frac{1}{\sqrt{2\pi}}\sum_{l,m}i^l\gamma_{lm}({\bf q})Y_{lm}^*(\theta , \zeta)\,\,,
\end{equation}
where $\theta$ is the angle between the incident electron direction in the laboratory frame and the body-frame dissociation axis {\bf R}, $\zeta$ 
is the azimuthal angle about ${\bf R}$ and $\gamma_{lm}({\bf q})$ is a partial resonance width which depends on the internal molecular coordinates denoted collectively by {\bf q} (see Fig.~\ref{thetafig}). By following the standard analysis for a narrow resonance~\cite{Taylor}, one can show that the partial widths are related to the total resonance width $\Gamma$ by the relation
\begin{equation}
\sum_{l,m}|\gamma_{lm}({\bf q})|^2= \Gamma({\bf q}).
\end{equation}
The resonance amplitudes $\gamma_{lm}$ thus suffice to completely determine the entrance amplitude $V_a$ for a narrow resonance. These amplitudes are in turn determined by fitting the calculated fixed-nuclei multi-channel $S$ matrix in the vicinity of the resonance energy to the generalized Breit-Wigner form
\begin{equation} 
\label{sres}
\begin{split}
S&=S^{bg} \times S^{res}\\
&=S^{bg} \times \left( 1 - \frac{iA}{E-E_R + i\Gamma/2} \right) ,
\end{split}
\end{equation}
where $S^{bg}$ is the background $S$ matrix and $A$ is
\begin{equation} 
A = S^{bg\dagger}B\,, 
\end{equation}
with 
\begin{equation}
\label{bmatrix}
B_{l m , \ l' m'} = \gamma_{l m} \gamma_{l' m'}.
\end{equation}
In the case of a polyatomic target, the orientation of the final neutral and/or ion fragments in the laboratory frame is generally not measured, so the attachment probability must be integrated over the angle $\zeta$ azimuthal to the recoil axis to produce an observable angular distribution:
\begin{equation} 
\label{DEA_diff}
\frac{d\sigma_{dea}}{d\theta}\propto \int d\zeta \left|\sum_{l,m}i^l\gamma_{lm}({\bf q})Y_{lm}^*(\theta , \zeta)\right|^2\,\,.
\end{equation} 

The quantity defined in Eq.~\ref{DEA_diff} coincides with the measured angular distribution if the overall rotation of the molecule is slow compared to dissociation and if the fragments recoil at a given angle in the molecular frame, i.e., if the molecular frame recoil axis ${\bf R}$ is constant over the Franck-Condon region of the neutral~\cite{Omalley68}.

\section{Results and Discussion}

\begin{figure}[ht]
\includegraphics[scale=0.55]{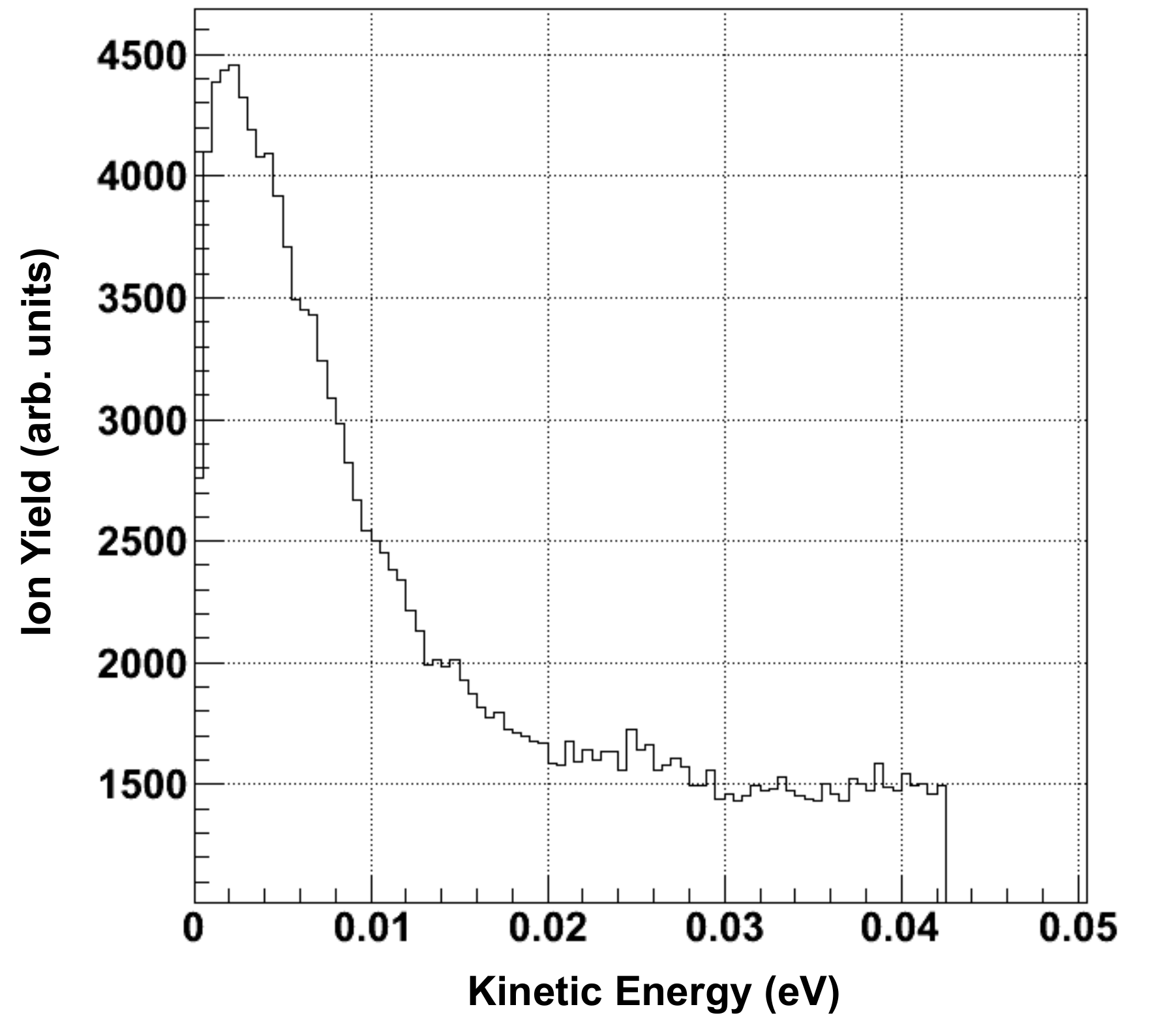}
\caption{C$_2$H$^-$ fragment kinetic energy release distribution for DEA at 3~eV.}
\label{kerplot}
\end{figure}



From the properly weighted momentum distribution in the right pane of Fig.~\ref{xtplot}, both the kinetic energy and fragment angular distributions can be investigated for DEA to the $^2\Pi_g$ resonance near 3~eV. Figure \ref{kerplot} shows the C$_2$H$^-$ fragment kinetic energy release (KER) distribution.  As can be seen in Fig.~\ref{kerplot}, the peak of the distribution reflects a small kinetic energy given to the C$_2$H$^-$ fragment during dissociation.  If we assume an electron energy of 3.0~eV, an electron affinity of 2.94~eV for C$_2$H \cite{janousek1979} and a C-H bond dissociation energy of 5.49~eV \cite{green1989}, the total amount of energy available during dissociation, neglecting internal excitation, is then 3.0 – (5.49 – 2.94) = 0.45~eV. The mass fraction of the C$_2$H$^-$ fragment then receives 1/26th of the total available kinetic energy which is 0.017~eV. In comparison to Fig.~\ref{kerplot}, the peak in the KER distribution, near 0.003~eV, is considerably less which could indicate some degree of internal excitation of the C$_2$H$^-$ fragment. We should note that the high-energy tail in Fig.~\ref{kerplot} is a result of the 0.5~eV resolution of the electron beam. If we substitute 3.5~eV for the electron energy in the previous calculation, the C$_2$H$^-$ fragment would receive a KER of 0.37~eV which corresponds to the flat region of the high-energy tail shown in Fig.~\ref{kerplot}. We also note that we have shown previously \cite{Moradmand13} that the weighting of the momentum distribution does not adversely affect the KER distribution. We have also determined the KER distribution by taking the full 4$\pi$ of the momentum data, i.e., without any weighting or slicing of the dissociation sphere, and arrive at the same KER distribution shown in Fig.~\ref{kerplot}.

\begin{figure}[ht]
\includegraphics[scale=0.8]{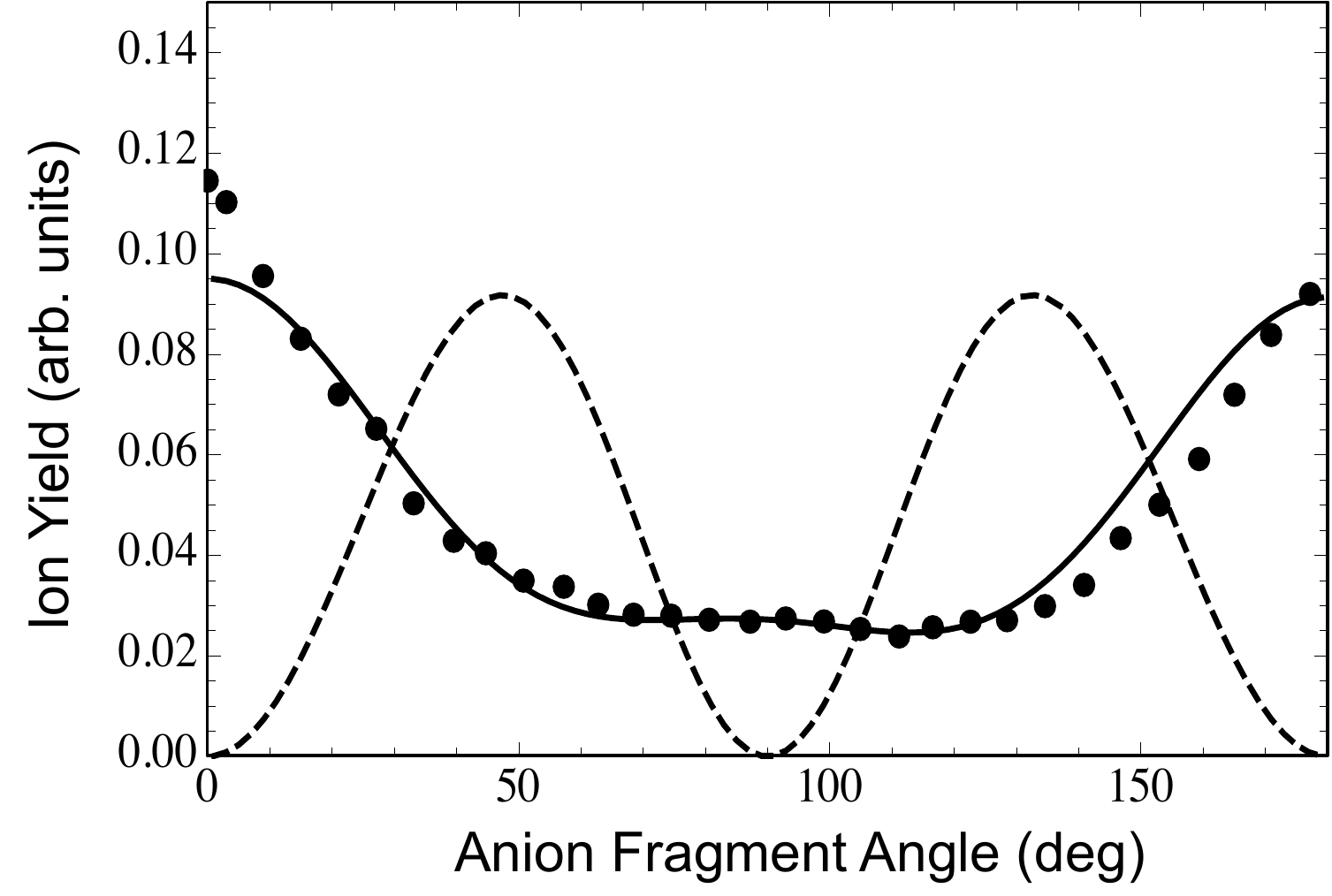}
\caption{C$_2$H$^-$ fragment angle distribution with respect to the incoming electron. The present data is shown by the circles. The dashed line represents the theoretically expected angular distribution without bending dynamics. The solid curve represents the theoretical nonaxial recoil dynamic for a bending angle of 27 degrees.}
\label{expangdist}
\end{figure}

Figure \ref{expangdist} shows the C$_2$H$^-$ fragment angular distribution with respect to the incoming electron. As can be seen in Figs.~\ref{xtplot} and \ref{expangdist}, the C$_2$H$^-$ fragments dissociated predominantly in the forward and backward directions with respect to the incoming electron. This is not reflective of the angular distribution expected from the attachment amplitude shown in Fig.~\ref{thetafig} for the case of axial recoil. The dashed line of Fig.~\ref{expangdist} illustrates the theoretically expected angular distribution of fragments in the case of axial recoil for the $^2\Pi_g$(A$'$) anion state .

In the present case, a simple axial recoil prediction cannot explain the measured data. Moreover, it is clear from the electronic structure calculations~\cite{Chourou} that for the HCCH$^-$ anion there is a large barrier to dissociation into HCC$^-$ + H in linear geometry, such that the H atom fragments only subsequent to C-C-H bending.  The barrier vanishes when the bending angle reaches $\sim$30 degrees. The dynamics therefore clearly violate the axial recoil approximation, in which the H atom is assumed to fragment along the linear molecular axis.

One way to deal with the problem of non-axial recoil would be to convolve the quantum mechanical attachment probability with recoil angle distributions obtained from classical trajectory calculations on the complex-valued anion potential energy surface~\cite{Adaniya09, Haxton11}. A simpler modification, which we have used in the present case, is to assume the H atom fragments at a single, but nonaxial, angle in the body frame~\cite{Moradmand13, Slaughter13}. We therefore first rotate the recoil axis by a fixed amount before averaging the attachment probability over the azimuthal angle $\zeta$ to obtain a fragment angular distribution (see Fig.~\ref{thetafig}).  Since the H fragment is light, its rotation angle is nearly equal to the C-C-H bending angle.

Since attachment can occur at geometries throughout the Franck-Condon region of the neutral and a forward/backward asymmetry is observed in the experiment, the attachment probabilities were computed and averaged over geometries corresponding to cis- and trans-bending, as well as asymmetric stretch, using two-point Hermite quadrature for each mode. We did not consider the symmetric stretch modes. We note that the attachment occurs preferentially to the stretched C-H bond when considering asymmetric stretch, so to model the observed forward/backward asymmetry we assume that dissociation occurs along the stretched C-H bond. The solid curve in Fig.~\ref{expangdist} reflects a C-C-H bending angle, and thus a $\theta$ rotation, of 27 degrees. This is in good agreement with the observed experimental fragment angular distribution, although the forward/backward asymmetry seen in the experiment is not completely reproduced by the theory.

\section{Conclusions}
We have performed ion-momentum imaging of the C$_2$H$^-$ fragments formed due to DEA of acetylene near the 3~eV $^2\Pi_g$ resonance. We observe a predominant anion fragment distribution in the forward and backward directions with respect to the incoming electron. This is at odds with the expected angular distribution that should result from the $^2\Pi_g$ anion state and indicates that a bending dynamics is present during dissociation. While previous experimental observations inferred a bending mechanism, there have been no direct observations. We have modeled this bending dynamic with \emph{ab initio} electronic structure and fixed-nuclei scattering calculations and achieve good agreement with the observed experimental angular fragment distribution. The observed KER distribution indicates that the anion fragments have low energy and that there is likely some internal excitation of the C$_2$H$^-$ fragments.

\begin{acknowledgments}
Work at University of California Lawrence Berkeley National Laboratory was performed under the auspices of the US Department of Energy under Contract DE-AC02-05CH11231 and was supported by the U.S. DOE Office of Basic Energy Sciences, Division of Chemical Sciences. Work at Auburn University was supported by the National Science Foundation under contract NSF-PHYS1404366. AEO acknowledges support by the National Science Foundation, with some of this material based on work while serving at NSF.
\end{acknowledgments}
\bibliography{HCCH_DEA}
\end{document}